\documentclass[a4paper]{jpconf}
\usepackage{graphicx}
\usepackage{amssymb}

\newtheorem{theorem}{Theorem}
\newtheorem{definition}{Definition}

\begin{document}

\title[Exotic $\mathbb{R}^{4}$ and QFT]{{\Large Exotic $\mathbb{R}^{4}$ and quantum field theory}{\normalsize }}
\author{{Torsten Asselmeyer-Maluga and Roland Mader}}
\address{German Aerospace center, Rutherfordstr. 2, 12489 Berlin, Germany }
\ead{torsten.asselmeyer-maluga@dlr.de}

\begin{abstract}
{\normalsize Recent work on exotic smooth $\mathbb{R}^{4}$'s, i.e.
topological $\mathbb{R}^{4}$ with exotic differential structure,
shows the connection of 4-exotics with the codimension-1 foliations
of $S^{3}$, $SU(2)$ WZW models and twisted K-theory $K_{H}(S^{3})$,
$H\in H^{3}(S^{3},\mathbb{Z})$. These results made it possible to
explicate some physical effects of exotic 4-smoothness. Here we present
a relation between exotic smooth $\mathbb{R}^{4}$ and operator algebras.
The correspondence uses the leaf space of the codimension-1 foliation
of $S^{3}$ inducing a von Neumann algebra $W(S^{3})$ as description.
This algebra is a type $I\! I\! I_{1}$ factor lying at the heart
of any observable algebra of QFT. By using the relation to factor
$I\! I$, we showed that the algebra $W(S^{3})$ can be interpreted
as Drinfeld-Turaev deformation quantization of the space of flat $SL(2,\mathbb{C})$
connections (or holonomies). Thus, we obtain a natural relation to
quantum field theory. Finally we discuss the appearance of concrete
action functionals for fermions or gauge fields and its connection
to quantum-field-theoretical models like the Tree QFT of Rivasseau.}{\normalsize \par}
\end{abstract}

%
%
%
%
%
%
%

\section{Introduction}

The construction of quantum theories from classical theories, known
as quantization, has a long and difficult history. It starts with
the discovery of quantum mechanics in 1925 and the formalization of
the quantization procedure by Dirac and von Neumann. The construction
of a quantum theory from a given classical one is highly non-trivial
and non-unique. But except for few examples, it is the only way which
will be gone today. From a physical point of view, the world surround
us is the result of an underlying quantum theory of its constituent
parts. So, one would expect that we must understand the transition
from the quantum to the classical world. But we had developed and
tested successfully the classical theories like mechanics or electrodynamics.
Therefore one tried to construct the quantum versions out of classical
theories. In this paper we will go the other way to obtain a quantum
field theory by geometrical methods and to show its equivalence to
a quantization of a classical Poisson algebra. 

The main technical tool will be the noncommutative geometry developed
by Connes \cite{Con:85}. Then intractable space like the leaf space
of a foliation can be described by noncommutative algebras. From the
physical point of view, we have now an interpretation of noncommutative
algebras (used in quantum theory) in a geometrical context. So, we
need only an idea for the suitable geometric structure. For that purpose
one formally considers the path integral over spacetime geometries.
In the evaluation of this integral, one has to include the possibility
of different smoothness structures for spacetime \cite{Pfeiffer2004,Ass2010}.
Brans \cite{BraRan:93,Bra:94a,Bra:94b} was the first who considered
exotic smoothness also on open smooth 4-manifolds as a possibility
for space-time. He conjectured that exotic smoothness induces an additional
gravitational field (\emph{Brans conjecture}). The conjecture was
established by Asselmeyer \cite{Ass:96} in the compact case and by
S\'ladkowski \cite{Sladkowski2001} in the non-compact case. S\'ladkowski
\cite{Sla:96,Sla:96b,Sla:96c} discussed the influence of differential
structures on the algebra $C(M)$ of functions over the manifold $M$
with methods known as non-commutative geometry. Especially in \cite{Sla:96b,Sla:96c}
he stated a remarkable connection between the spectra of differential
operators and differential structures. But there is a big problem
which prevents progress in the understanding of exotic smoothness
especially for the $\mathbb{R}^{4}$: there is no known explicit coordinate
representation. As the result no exotic smooth function on any such
$\mathbb{R}^{4}$ is known even though there exist families of infinite
continuum many different non diffeomorphic smooth $\mathbb{R}^{4}$.
This is also a strong limitation for the applicability to physics
of non-standard open 4-smoothness. Bizaca \cite{Biz:94} was able
to construct an infinite coordinate patch by using Casson handles.
But it still seems hopeless to extract physical information from that
approach.

This situation is not satisfactory but we found a possible solution.
The solution is a careful analysis of the small exotic $\mathbb{R}^{4}$
by using foliation theory (see especially \cite{AsselmeyerKrol2009,AsselmKrol2011c})
to derive a relation between exotic smoothness and codimension-1 foliations
(see Theorem \ref{thm:codim-1-foli-radial-fam}). By using noncommutative
geometry, this approach is able to produce a von Neumann algebra via
the leaf space of the foliation which can be interpreted as the observable
algebra of some QFT (see \cite{Haag:96}). Fortunately, our approach
to exotic smoothness is strongly connected with a codimension-1 foliation
of type $I\! I\! I$, i.e. the leaf space is a factor $I\! I\! I_{1}$
von Neumann algebra. Especially this algebra is the preferred algebra
in the local algebra approach to QFT \cite{Haag:96,Borchers2000}.
Recently, this factor $I\! I\! I$ case was also discussed in connection
with quantum gravity (via the spectral triple of Connes) \cite{BertozziniConti2010}. 

In this paper we present a deep connection between an exotic $\mathbb{R}^{4}$
as variation of the usual Minkowski space with $\mathbb{R}^{4}$-topology.
We start with some introductionary material in the next section. In
section \ref{sec:Quantization} we present the main result about a
relation between exotic $\mathbb{R}^{4}$ and type $I\! I\! I_{1}$
factor von Neumann algebras leading to a possible relation to Quantum
field theory (QFT). This relation is presented in the last section
with some remarks about the construction of the action functional.

\section{From exotic $\mathbb{R}^{4}$ to operators algebras}

Einsteins insight that gravity is the manifestation of geometry leads
to a new view on the structure of spacetime. From the mathematical
point of view, spacetime is a smooth 4-manifold endowed with a (smooth)
metric as basic variable for general relativity. Later on, the existence
question for Lorentz structure and causality problems (see Hawking
and Ellis \cite{HawEll:94}) gave further restrictions on the 4-manifold:
causality implies non-compactness, Lorentz structure needs a codimension-1
foliation. Usually, one starts with a globally foliated, non-compact
4-manifold $\Sigma\times\mathbb{R}$ fulfilling all restrictions where
$\Sigma$ is a smooth 3-manifold representing the spatial part. But
other non-compact 4-manifolds are also possible, i.e. it is enough
to assume a non-compact, smooth 4-manifold endowed with a codimension-1
foliation.

All these restrictions on the representation of spacetime by the manifold
concept are clearly motivated by physical questions. Among the properties
there is one distinguished element: the smoothness. Usually one assumes
a smooth, unique atlas of charts (i.e. a smooth or differential structure)
covering the manifold where the smoothness is induced by the unique
smooth structure on $\mathbb{R}$. But that is not the full story.
Even in dimension 4, there are an infinity of possible other smoothness
structures (i.e. a smooth atlas) non-diffeomorphic to each other.
For a deeper insight we refer to the book \cite{Asselmeyer2007}.

\subsection{Exotic 4-manifolds and exotic $\mathbb{R}^{4}$}

If two manifolds are homeomorphic but non-diffeomorphic, they are
\textbf{exotic} to each other. The smoothness structure is called
an \textbf{exotic smoothness structure}.

The implications for physics are tremendous because we rely on the
smooth calculus to formulate field theories. Thus different smoothness
structures have to represent different physical situations leading
to different measurable results. But it should be stressed that \emph{exotic
smoothness is not exotic physics.} Exotic smoothness is part of topology
(differential topology), i.e. a finer determination of the topology
(the smooth atlas of the manifold) fulfilling the condition of smoothness.
Therefore we obtain another parameter to vary a manifold
with fixed topology. Usually one starts with a topological manifold
$M$ and introduces structures on them. Then one has the following
ladder of possible structures:\begin{eqnarray*}
\mbox{Topology}\to & \mbox{\mbox{piecewise-linear(PL)}}\to & \mbox{Smoothness}\\
\to & \mbox{bundles, Lorentz, Spin etc.}\to & \mbox{metric,...}\end{eqnarray*}
 We do not want to discuss the first transition, i.e. the existence
of a triangulation on a topological manifold. But we remark that the
existence of a PL structure implies uniquely a smoothness structure
in all dimensions smaller than 7 \cite{KirSie:77}. 

Given two homeomorphic manifolds $M,M'$, how can we compare both
manifolds to decide whether both are diffeomorphic? A mapping between
two manifolds $M,M'$ can be described by a $(n+1)-$dimensional manifold
$W$ with $\partial W=M\sqcup M'$, called a cobordism $W$. In the
following, the two homeomorphic manifolds $M,M'$ are simple-connected.
The celebrated h-cobordism theorem of Smale \cite{Sma:61,Sma:62}
gives a simple criteria when a diffeomorphism between $M$ and $M'$
for dimension greater than $5$ exists: there is a cobordism $W$
between $M$ and $M'$ where the inclusions $M,M'\hookrightarrow W$
induce homotopy-equivalences between $M,M'$ and $W$. We call $W$
a h-cobordism. Therefore the classification problem of smoothness
structures in higher dimensions ($>5$) is a homotopy-theoretic problem.
The set of smooth structures (up to h-cobordisms) of the $n-$sphere
$S^{n}$ has the structure of an abelian group $\Theta_{n}$ (under
connected sum $\#$). Via the h-cobordism theorem this group is identical
to the group of homotopy spheres which was analyzed by Kervaire and
Milnor \cite{KerMil:63} to be a finite group. Later on the result
about the exotic spheres can be extended to any smoothable manifold
\cite{KirSie:77}, i.e. there is only a finite number of non-diffeomorphic
smoothness structures. In all dimensions smaller than four, there
is an unique smoothness structure, the standard structure. But all
methods failed for the special case of dimension four.

Now we consider two homeomorphic, smooth, but non-diffeomorphic 4-manifolds
$M_{0}$ and $M$. As expressed above, a comparison of both smoothness
structures is given by a h-cobordism $W$ between $M_{0}$ and $M$
($M,M_{0}$ are homeomorphic). Let the 4-manifolds additionally be
compact, closed and simple-connected, then we have the structure theorem%
\footnote{A diffeomorphism will be described by the symbol $=$ in the following.%
} of h-cobordisms \cite{CuFrHsSt:97}: \begin{theorem} \label{thm:Akbulut-cork}Let
$W$ be a h-cobordisms between $M_{0},\, M$, then there are contractable
submanifolds $A_{0}\subset M_{0},\,\, A\subset M$ and a h subcobordism
$X\subset W$ with $\partial X=A_{0}\sqcup A$, so that the remaining
h-cobordism $W\setminus X$ trivializes $W\setminus X=(M_{0}\setminus A_{0})\times[0,1]$
inducing a diffeomorphism between $M_{0}\setminus A_{0}$ and $M\setminus A$.
\end{theorem} In short that means that the smoothness structure of
$M$ is determined by the contractable manifold $A$ -- \emph{its}
\emph{Akbulut cork} -- and by the embedding of $A$ into $M$. As
shown by Freedman\cite{Fre:82}, the Akbulut cork has a homology 3-sphere%
\footnote{A homology 3-sphere is a 3-manifold with the same homology as the
3-sphere $S^{3}$.%
} as boundary. The embedding of the cork can be derived now from the
structure of the h-subcobordism $X$ between $A_{0}$ and $A$. For
that purpose we cut $A_{0}$ out from $M_{0}$ and $A$ out from $M$.
Then we glue in both submanifolds $A_{0},A$ via the maps $\tau_{0}:\partial A_{0}\to\partial(M_{0}\setminus A_{0})=\partial A_{0}$
and $\tau:\partial A\to\partial(M\setminus A)=\partial A$. Both maps
$\tau_{0},\tau$ are involutions, i.e. $\tau\circ\tau=id$. One of
these maps (say $\tau_{0}$) can be chosen to be trivial (say $\tau_{0}=id$).
Thus the involution $\tau$ determines the smoothness structure. Especially
the topology of the Akbulut cork $A$ and its boundary $\partial A$
is given by the topology of $M$. For instance, the Akbulut cork of
the blow-uped 4-dimensional \emph{K3 surface $K3\#\overline{\mathbb{C}P}^{2}$}
is the so-called \emph{Mazur manifold} \cite{AkbKir:79,Akb:91} with
the \emph{Brieskorn-Sphere} $B(2,5,7)$ as boundary. Akbulut and its
coworkers \cite{Akbulut08,Akbulut09} discuss many examples of Akbulut
corks and the dependence of the smoothness structure on the cork.

Then as shown by Bizaca and Gompf \cite{BizGom:96} the neighborhood
$N$ of the handle pair as well the neighborhood $N(A)$ of the embedded
Akbulut cork consists of the cork $A$ and the Casson handle $CH$.
Especially the open neighborhood $N(A)$ of the Akbulut cork is an
exotic $\mathbb{R}^{4}$. The situation was analyzed in \cite{GomSti:1999}:
\begin{theorem} \label{thm:fail-h-cobordism-exotic-R4}Let $W^{5}$
be a non-trivial (smooth) h-cobordism between $M_{0}^{4}$ and $M^{4}$
(i.e. $W$ is not diffeomorphic to $M\times[0,1]$). Then there is
an open sub-h-cobordism $U^{5}$ that is homeomorphic to $\mathbb{R}^{4}\times[0,1]$
and contains a compact contractable sub-h-cobordism $X$ (the cobordism
between the Akbulut corks, see above), such that both $W$ and $U$
are trivial cobordisms outside of $X$, i.e. one has the diffeomorphisms
\begin{eqnarray*}
W\setminus X=((W\cap M)\setminus X)\times[0,1] & \quad\mbox{and}\\
\quad U\setminus X=((U\cap M)\setminus X)\times[0,1]\end{eqnarray*}
 (the latter can be chosen to be the restriction of the former). Furthermore
the open sets $U\cap M$ and $U\cap M_{0}$ are homeomorphic to $\mathbb{R}^{4}$
which are exotic $\mathbb{R}^{4}$ if $W$ is non-trivial. \end{theorem}
Then one gets an exotic $\mathbb{R}^{4}$ which smoothly embeds automatically
in the 4-sphere, called a small exotic $\mathbb{R}^{4}$. Furthermore
we remark that the exoticness of the $\mathbb{R}^{4}$ is connected
with the non-trivial smooth h-cobordism $W^{5}$, i.e. the failure
of the smooth h-cobordism theorem implies the existence of small exotic
$\mathbb{R}^{4}$'s.

Let $R$ be a small exotic $\mathbb{R}^{4}$ induced from the non-product
h-cobordism $W$ between $M$ and $M_{0}$ with Akbulut corks $A\subset M$
and $A_{0}\subset M_{0}$, respectively. Let $K\subset\mathbb{R}^{4}$
be a compact subset. Bizaca and Gompf \cite{BizGom:96} constructed
the small exotic $\mathbb{R}_{1}^{4}$ by using the simplest tree
$Tree_{+}$. Bizaca \cite{Biz:94,Bizaca1995} showed that the Casson
handle generated by $Tree_{+}$ is an exotic Casson handle. Using
Theorem 3.2 of \cite{DeMichFreedman1992}, there is a topological
radius function $\rho:\mathbb{R}_{1}^{4}\to[0,+\infty)$ (polar coordinates)
so that $\mathbb{R}_{t}^{4}=\rho^{-1}([0,r))$ with $t=1-\frac{1}{r}$.
Then $K\subset\mathbb{R}_{0}^{4}$ and $\mathbb{R}_{t}^{4}$ is also
a small exotic $\mathbb{R}^{4}$ for $t$ belonging to a Cantor set
$CS\subset[0,1]$. Especially two exotic $\mathbb{R}_{s}^{4}$ and
$\mathbb{R}_{t}^{4}$ are non-diffeomorphic for $s<t$ except for
countable many pairs. In \cite{DeMichFreedman1992} it was claimed
that there is a smoothly embedded homology 4-disk $A$. The boundary
$\partial A$ is a homology 3-sphere with a non-trivial representation
of its fundamental group into $SO(3)$ (so $\partial A$ cannot be
diffeomorphic to a 3-sphere). According to Theorem \ref{thm:fail-h-cobordism-exotic-R4}
this homology 4-disk must be identified with the Akbulut cork of the
non-trivial h-cobordism. The cork $A$ is contractable and can be
(at least) build by one 1-handle and one 2-handle (case of a Mazur
manifold). Given a radial family $\mathbb{R}_{t}^{4}$ with radius
$r=\frac{1}{1-t}$ so that $t=1-\frac{1}{r}\subset CS\subset[0,1]$.
Suppose there is a diffeomorphism\[
(d,id_{K}):(\mathbb{R}_{s}^{4},K)\to(\mathbb{R}_{t}^{4},K)\qquad s\not=t\in CS\]
 fixing the compact subset $K$. Then this map $d$ induces end-periodic
manifolds%
\footnote{We ignore the inclusion for simplicity.%
} $M\setminus(\bigcap_{i=0}^{\infty}d^{i}(\mathbb{R}_{s}^{4}))$ and
$M_{0}\setminus(\bigcap_{i=0}^{\infty}d^{i}(\mathbb{R}_{s}^{4}))$
which must be smoothable contradicting a theorem of Taubes \cite{Tau:87}.
Therefore $\mathbb{R}_{s}^{4}$ and $\mathbb{R}_{t}^{4}$ are non-diffeomorphic
for $t\not=s$ (except for countable many possibilities).

\subsection{Exotic $\mathbb{R}^{4}$ and codimension-1 foliations}

In this subsection we will construct a codimension-one foliation on
the boundary $\partial A$ of the cork with non-trivial Godbillon-Vey
invariant. 

A codimension $k$ foliation%
\footnote{In general, the differentiability of a foliation is very important.
Here we consider the smooth case only. %
} of an $n$-manifold $M^{n}$ (see the nice overview article \cite{Law:74})
is a geometric structure which is formally defined by an atlas $\left\{ \phi_{i}:U_{i}\to M^{n}\right\} $,
with $U_{i}\subset\mathbb{R}^{n-k}\times\mathbb{R}^{k}$, such that
the transition functions have the form \[
\phi_{ij}(x,y)=(f(x,y),g(y)),\,\left[x\in\mathbb{R}^{n-k},y\in\mathbb{R}^{k}\right]\quad.\]
Intuitively, a foliation is a pattern of $(n-k)$-dimensional stripes
- i.e., submanifolds - on $M^{n}$, called the leaves of the foliation,
which are locally well-behaved. The tangent space to the leaves of
a foliation $\mathcal{F}$ forms a vector bundle over $M^{n}$, denoted
$T\mathcal{F}$. The complementary bundle $\nu\mathcal{F}=TM^{n}/T\mathcal{F}$
is the normal bundle of $\mathcal{F}$. Such foliations are called
regular in contrast to singular foliations or Haefliger structures.
For the important case of a codimension-1 foliation we need an overall
non-vanishing vector field or its dual, an one-form $\omega$. This
one-form defines a foliation iff it is integrable, i.e.\begin{equation}
d\omega\wedge\omega=0\label{eq:integrability-foliation}\end{equation}
The cross-product $M\times N$ defines for example a trivial foliation.
Now we will discuss an important equivalence relation between foliations,
cobordant foliations. \begin{definition} Let $M_{0}$ and $M_{1}$
be two closed, oriented $m$-manifolds with codimension-$q$ foliations.
Then these foliated manifolds are said to be \emph{foliated cobordant}
if there is a compact, oriented $(m+1)$-manifold with boundary $\partial W=M_{0}\sqcup\overline{M}_{1}$
and with a codimension-$q$ foliation $\mathcal{F}$ transverse to
the boundary. Every leaf $L_{\alpha}$ of the foliation $\mathcal{F}$
induces leafs $L_{\alpha}\cap\partial W$ of foliations $\mathcal{F}_{M_{0}},\mathcal{F}_{M_{1}}$on
the two components of the boundary $\partial W$.\end{definition}
The resulted foliated cobordism classes $[\mathcal{F}_{M}]$ of the
manifold $M$ form an abelian group $\mathcal{CF}_{m,q}(M)$ under
disjoint union $\sqcup$(inverse $\overline{M}$, unit $S^{q}\times S^{m-q}$,
see \cite{Tamura1992} \S29). 

In \cite{Thu:72}, Thurston constructed a foliation of the 3-sphere
$S^{3}$ depending on a polygon $P$ in the hyperbolic plane $\mathbb{H}^{2}$
so that two foliations are non-cobordant if the corresponding polygons
have different areas. Now we consider two codimension-1 foliations
$\mathcal{F}_{1},\mathcal{F}_{2}$ depending on the convex polygons
$P_{1}$ and $P_{2}$ in $\mathbb{H}^{2}$. As mentioned above, these
foliations $\mathcal{F}_{1},\mathcal{F}_{2}$ are defined by two one-forms
$\omega_{1}$ and $\omega_{2}$ with $d\omega_{a}\wedge\omega_{a}=0$
and $a=0,1$. Now we define the one-forms $\theta_{a}$ as the solution
of the equation\begin{equation}
d\omega_{a}=-\theta_{a}\wedge\omega_{a}\label{eq:solution-Godbillon-Vey}\end{equation}
 and consider the closed 3-form\begin{equation}
\Gamma_{\mathcal{F}_{a}}=\theta_{a}\wedge d\theta_{a}\label{eq:Godbillon-Vey-class}\end{equation}
associated to the foliation $\mathcal{F}_{a}$. As discovered by Godbillon
and Vey \cite{GodVey:71}, $\Gamma_{\mathcal{F}}$ depends only on
the foliation $\mathcal{F}$ and not on the realization via $\omega,\theta$.
Thus $\Gamma_{\mathcal{F}}$, the \emph{Godbillon-Vey class}, is an
invariant of the foliation. Let $\mathcal{F}_{1}$ and $\mathcal{F}_{2}$
be two cobordant foliations then $\Gamma_{\mathcal{F}_{1}}=\Gamma_{\mathcal{F}_{2}}$.
Thurston \cite{Thu:72} obtains for the Godbillon-Vey number \begin{equation}
GV(S^{3},\mathcal{F}_{a})=\langle\Gamma_{\mathcal{F}_{a}},[S^{3}]\rangle=\intop_{S^{3}}\Gamma_{\mathcal{F}_{a}}=vol(\pi^{-1}(Q))=4\pi\cdot Area(P_{a})\label{eq:GV-number-Thurston-foliation}\end{equation}
where only the foliation on $M=(S^{2}\setminus\left\{ \mbox{\mbox{k} punctures}\right\} )\times S^{1}$
contributes to the Godbillon-Vey class because the Reeb foliations
have vanishing Godbillon-Vey class. Let $[1]\in H^{3}(S^{3},\mathbb{R})$
be the dual of the fundamental class $[S^{3}]$ defined by the volume
form, then the Godbillon-Vey class can be represented by\begin{equation}
\Gamma_{\mathcal{F}_{a}}=4\pi\cdot Area(P_{a})[1]\label{eq:Godbillon-Vey-class-Thurston-foliation}\end{equation}
Furthermore the Godbillon-Vey number $GV$ seen as linear functional
defines a surjective homomorphism \begin{equation}
GV:\mathcal{CF}_{3,1}(S^{3})\to\mathbb{R}\label{eq:GV-homomorphism}\end{equation}
from the group of foliated cobordisms $\mathcal{CF}_{3,1}(S^{3})$
of the 3-sphere to the real numbers of possible areas $Area(P)$,
i.e.
\begin{itemize}
\item $\mathcal{F}_{1}$ is cobordant to $\mathcal{F}_{2}$ $\Longrightarrow$$Area(P_{1})=Area(P_{2})$
(the reverse direction (injectivity) is open)
\item $\mathcal{F}_{1}$ and $\mathcal{F}_{2}$ are non-cobordant $\Longleftrightarrow$$Area(P_{1})\not=Area(P_{2})$ 
\end{itemize}
We note that $Area(P)=(k-2)\pi-\sum_{k}\alpha_{k}$. The Godbillon-Vey
class is an element of the deRham cohomology $H^{3}(S^{3},\mathbb{R})$
which will be used later to construct a relation to gerbes. Furthermore
we remark that the classification is not complete. Thurston constructed
only a surjective homomorphism from the group of cobordism classes
of foliation of $S^{3}$ into the real numbers $\mathbb{R}$. We remark
the close connection between the Godbillon-Vey class (\ref{eq:Godbillon-Vey-class})
and the Chern-Simons form if $\theta$ can be interpreted as connection
of a suitable line bundle.

In \cite{AsselmeyerKrol2009} we presented the relation between foliations
on the boundary $\partial A$ of the cork $A$ and the radial family
o small exotic $\mathbb{R}^{4}$. Especially we have: \begin{theorem}
\label{thm:codim-1-foli-radial-fam}Given a radial family $R_{t}$
of small exotic $\mathbb{R}_{t}^{4}$ with radius $r$ and $t=1-\frac{1}{r}\subset CS\subset[0,1]$
induced from the non-product h-cobordism $W$ between $M$ and $M_{0}$
with Akbulut cork $A\subset M$ and $A\subset M_{0}$, respectively.
The radial family $R_{t}$ determines a family of codimension-one
foliations of $\partial A$ with Godbillon-Vey number $r^{2}$. Furthermore
given two exotic spaces $R_{t}$ and $R_{s}$, homeomorphic but non-diffeomorphic
to each other (and so $t\not=s$) then the two corresponding codimension-one
foliation of $\partial A$ are non-cobordant to each other.\end{theorem}

\subsection{From exotic smoothness to operator algebras\label{sub:From-exotic-smoothness}}

\begin{figure}
\begin{center}
\includegraphics[scale=0.25]{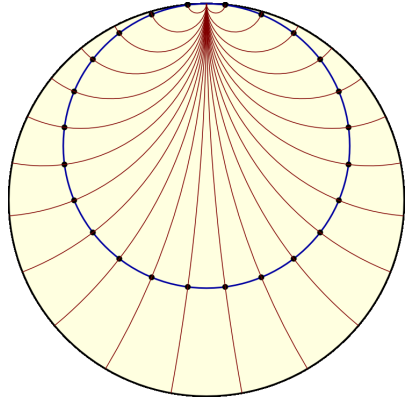}
\end{center}
\caption{horocycle, a curve whose normals all converge asymptotically\label{fig:horocycle-fig}}
\end{figure}

Connes \cite{Connes94} constructed the operator algebra $C_{r}^{*}(M,F)$
associated to a foliation $F$. The correspondence between a foliation
and the operator algebra (as well as the von Neumann algebra) is visualized
by table \ref{tab:relation-foliation-operator}. %
\begin{table}
\caption{relation between foliation and operator algebra\label{tab:relation-foliation-operator}}
\begin{center}
\begin{tabular}{|c|c|}
\hline 
Foliation & Operator algebra\tabularnewline
\hline
\hline 
leaf & operator\tabularnewline
\hline 
closed curve transversal to foliation & projector (idempotent operator)\tabularnewline
\hline 
holonomy & linear functional (state)\tabularnewline
\hline 
local chart & center of algebra\tabularnewline
\hline
\end{tabular}
\end{center}
\end{table}
 As extract of our previous paper \cite{AsselmeyerKrol2009}, we obtained
a relation between exotic $\mathbb{R}^{4}$'s and codimension-1 foliations
of the 3-sphere $S^{3}$. For a codimension-1 foliation there is the
Godbillon-Vey invariant \cite{GodVey:71} as element of $H^{3}(M,\mathbb{R})$.
Hurder and Katok \cite{HurKat:84} showed that the $C^{*}$algebra
of a foliation with non-trivial Godbillon-Vey invariant contains a
factor $I\! I\! I$ subalgebra. Using Tomita-Takesaki-theory, one
has a continuous decomposition (as crossed product) of any factor
$I\! I\! I$ algebra $M$ into a factor $I\! I_{\infty}$ algebra
$N$ together with a one-parameter group%
\footnote{The group $\mathbb{R}_{+}^{*}$ is the group of positive real numbers
with multiplication as group operation also known as Pontrjagin dual.%
} $\left(\theta_{\lambda}\right)_{\lambda\in\mathbb{R}_{+}^{*}}$ of
automorphisms $\theta_{\lambda}\in Aut(N)$ of $N$, i.e. one obtains

\[
M=N\rtimes_{\theta}\mathbb{R}_{+}^{*}\quad.\]
But that means, there is a foliation induced from the foliation of
the $S^{3}$ producing this $I\! I_{\infty}$ factor. Connes \cite{Connes94}
(in section I.4 page 57ff) constructed the foliation $F'$ canonically
associated to $F$ having the factor $I\! I_{\infty}$ as von Neumann
algebra. In our case it is the horocycle flow: Let $P$ the polygon
on the hyperbolic space $\mathbb{H}^{2}$ determining the foliation
of the $S^{3}$ (see above). $P$ is equipped with the hyperbolic
metric $2|dz|/(1-|z|^{2})$ together with the collection $T_{1}P$
of unit tangent vectors to $P$. A horocycle in $P$ is a circle contained
in $P$ which touches $\partial P$ at one point (see Fig. \ref{fig:horocycle-fig}).
%
%
 Then the horocycle flow $T_{1}P\to T_{1}P$ is the flow moving an
unit tangent vector along a horocycle (in positive direction at unit
speed). Then the polygon $P$ determines a surface $S$ of genus $g>1$
with abelian torsion-less fundamental group $\pi_{1}(S)$ so that
the homomorphism $\pi_{1}(S)\to\mathbb{R}$ determines an unique (ergodic
invariant) Radon measure. Finally the horocycle flow determines a
factor $I\! I_{\infty}$ foliation associated to the factor $I\! I\! I_{1}$
foliation. We remark for later usage that this foliation is determined
by a set of closed curves (the horocycles). Furthermore the idempotent
operators in the operator algebra $C_{r}^{*}(M,F)$ of the foliation
$F$ are represented by closed curves transversal to the foliation
(see \cite{Connes94}).

\section{Quantization\label{sec:Quantization}}

In this section we describe a deep relation between quantization and
the codimension-1 foliation of the $S^{3}$ determining the smoothness
structure on a small exotic $\mathbb{R}^{4}$. Here and in the following
\emph{we will identify the leaf space with its operator algebra}.

\subsection{The observable algebra and Poisson structure\label{sub:The-observable-algebra}}

In this section we will describe the formal structure of a classical
theory coming from the algebra of observables using the concept of
a Poisson algebra. In quantum theory, an observable is represented
by a hermitean operator having the spectral decomposition via projectors
or idempotent operators. The coefficient of the projector is the eigenvalue
of the observable or one possible result of a measurement. At least
one of these projectors represent (via the GNS representation) a quasi-classical
state. Thus to construct the substitute of a classical observable
algebra with Poisson algebra structure we have to concentrate on the
idempotents in the $C^{*}$ algebra. Now we will see that the set
of closed curves on a surface has the structure of a Poisson algebra.
Let us start with the definition of a Poisson algebra. \begin{definition} 

Let $P$ be a commutative algebra with unit over $\mathbb{R}$ or
$\mathbb{C}$. A \emph{Poisson bracket} on $P$ is a bilinearform
$\left\{ \:,\:\right\} :P\otimes P\to P$ fulfilling the following
3 conditions:

anti-symmetry $\left\{ a,b\right\} =-\left\{ b,a\right\} $

Jacobi identity $\left\{ a,\left\{ b,c\right\} \right\} +\left\{ c,\left\{ a,b\right\} \right\} +\left\{ b,\left\{ c,a\right\} \right\} =0$

derivation $\left\{ ab,c\right\} =a\left\{ b,c\right\} +b\left\{ a,c\right\} $.\\
Then a \emph{Poisson algebra} is the algebra $(P,\{\,,\,\})$.
\end{definition}

Now we consider a surface $S$ together with a closed curve $\gamma$.
Additionally we have a Lie group $G$ given by the isometry group.
The closed curve is one element of the fundamental group $\pi_{1}(S)$.
From the theory of surfaces we know that $\pi_{1}(S)$ is a free abelian
group. Denote by $Z$ the free $\mathbb{K}$-module ($\mathbb{K}$
a ring with unit) with the basis $\pi_{1}(S)$, i.e. $Z$ is a freely
generated $\mathbb{K}$-modul. Recall Goldman's definition of the
Lie bracket in $Z$ (see \cite{Goldman1984}). For a loop $\gamma:S^{1}\to S$
we denote its class in $\pi_{1}(S)$ by $\left\langle \gamma\right\rangle $.
Let $\alpha,\beta$ be two loops on $S$ lying in general position.
Denote the (finite) set $\alpha(S^{1})\cap\beta(S^{1})$ by $\alpha\#\beta$.
For $q\in\alpha\#\beta$ denote by $\epsilon(q;\alpha,\beta)=\pm1$
the intersection index of $\alpha$ and $\beta$ in $q$. Denote by
$\alpha_{q}\beta_{q}$ the product of the loops $\alpha,\beta$ based
in $q$. Up to homotopy the loop $(\alpha_{q}\beta_{q})(S^{1})$ is
obtained from $\alpha(S^{1})\cup\beta(S^{1})$ by the orientation
preserving smoothing of the crossing in the point $q$. Set \begin{equation}
[\left\langle \alpha\right\rangle ,\left\langle \beta\right\rangle ]=\sum_{q\in\alpha\#\beta}\epsilon(q;\alpha,\beta)(\alpha_{q}\beta_{q})\quad.\label{eq:Lie-bracket-loops}\end{equation}
According to Goldman \cite{Goldman1984}, Theorem 5.2, the bilinear
pairing $[\,,\,]:Z\times Z\to Z$ given by (\ref{eq:Lie-bracket-loops})
on the generators is well defined and makes $Z$ to a Lie algebra.
The algebra $Sym(Z)$ of symmetric tensors is then a Poisson algebra
(see Turaev \cite{Turaev1991}).

Now we introduce a principal $G$ bundle on $S$, representing a geometry
on the surface. This bundle is induced from a $G$ bundle over $S\times[0,1]$
having always a flat connection. Alternatively one can consider a
homomorphism $\pi_{1}(S)\to G$ represented as holonomy functional\begin{equation}
hol(\omega,\gamma)=\mathcal{P}\exp\left(\int\limits _{\gamma}\omega\right)\in G\label{eq:holonomy-definition}\end{equation}
with the path ordering operator $\mathcal{P}$ and $\omega$ as flat
connection (i.e. inducing a flat curvature $\Omega=d\omega+\omega\wedge\omega=0$).
This functional is unique up to conjugation induced by a gauge transformation
of the connection. Thus we have to consider the conjugation classes
of maps\[
hol:\pi_{1}(S)\to G\]
forming the space $X(S,G)$ of gauge-invariant flat connections of
principal $G$ bundles over $S$. Then \cite{Skovborg2006} constructed
of the Poisson structure on , i.e. the space $X(S,G)$ has a natural
Poisson structure (induced by the bilinear form (\ref{eq:Lie-bracket-loops})
on the group) and the Poisson algebra \emph{$(X(S,G),\left\{ \,,\,\right\} )$}
of complex functions over them is the algebra of observables. For
the following we will fix the group to be $G=SL(2,\mathbb{C})$ as
the largest isometry group of a homogeneous 3-manifold (or space of
constant curvature). The space $X(S,SL(2,\mathbb{C}))$ has a natural
Poisson structure (induced by the bilinear form of Goldman \cite{Goldman1984}
on the group) and the Poisson algebra \emph{$(X(S,SL(2,\mathbb{C}),\left\{ \,,\,\right\} )$}
of complex functions over them is the algebra of observables.

\subsection{Drinfeld-Turaev Quantization\label{sub:Drinfeld-Turaev-Quantization}}

Now we introduce the ring $\mathbb{C}[[h]]$ of formal polynomials
in $h$ with values in $\mathbb{C}$. This ring has a topological
structure, i.e. for a given power series $a\in\mathbb{C}[[h]]$ the
set $a+h^{n}\mathbb{C}[[h]]$ forms a neighborhood. Now we define 

\begin{definition} A \emph{Quantization} of a Poisson algebra $P$
is a $\mathbb{C}[[h]]$ algebra $P_{h}$ together with the $\mathbb{C}$-algebra
isomorphism $\Theta:P_{h}/hP\to P$ so that

1. the modul $P_{h}$ is isomorphic to $V[[h]]$ for a $\mathbb{C}$
vector space $V$

2. let $a,b\in P$ and $a',b'\in P_{h}$ be $\Theta(a)=a'$, $\Theta(b)=b'$
then\[
\Theta\left(\frac{a'b'-b'a'}{h}\right)=\left\{ a,b\right\} \]
\end{definition}

One speaks of a deformation of the Poisson algebra by using a deformation
parameter $h$ to get a relation between the Poisson bracket and the
commutator. Therefore we have the problem to find the deformation
of the Poisson algebra $(X(S,SL(2,\mathbb{C})),\left\{ \,,\,\right\} )$.
The solution to this problem can be found via two steps: 
\begin{enumerate}
\item at first find another description of the Poisson algebra by a structure
with one parameter at a special value and 
\item secondly vary this parameter to get the deformation. 
\end{enumerate}
Fortunately both problems were already solved (see \cite{Turaev1989,Turaev1991}).
The solution of the first problem is : 

\emph{The Skein modul $K_{-1}(S\times[0,1])$ (i.e. $t=-1$) has the
structure of an algebra isomorphic to the Poisson algebra $(X(S,SL(2,\mathbb{C}),\left\{ \,,\,\right\} )$.}
\emph{(see \cite{BulPrzy:1999,Bullock1999}) }

Then we have also the solution of the second problem: 

\emph{The skein algebra $K_{t}(S\times[0,1])$ is the quantization
of the Poisson algebra $(X(S,SL(2,\mathbb{C}),\left\{ \,,\,\right\} )$
with the deformation parameter $t=\exp(h/4)$.(see \cite{BulPrzy:1999})} 

\begin{figure}
\begin{center}\includegraphics[scale=0.25]{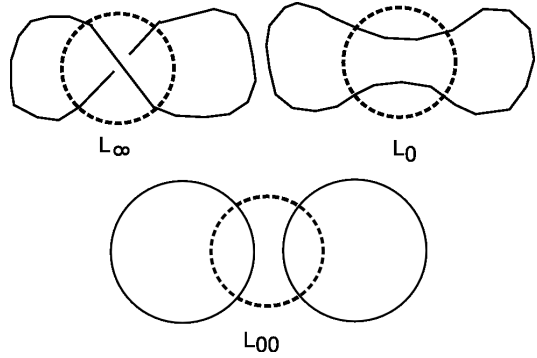}\end{center}

\caption{crossings $L_{\infty},L_{o},L_{oo}$\label{fig:skein-crossings}}

\end{figure}

To understand these solutions we have to introduce the skein module
$K_{t}(M)$ of a 3-manifold $M$ (see \cite{PrasSoss:97}). For that
purpose we consider the set of links $\mathcal{L}(M)$ in $M$ up
to isotopy and construct the vector space $\mathbb{C}\mathcal{L}(M)$
with basis $\mathcal{L}(M)$. Then one can define $\mathbb{C}\mathcal{L}[[t]]$
as ring of formal polynomials having coefficients in $\mathbb{C}\mathcal{L}(M)$.
Now we consider the link diagram of a link, i.e. the projection of
the link to the $\mathbb{R}^{2}$ having the crossings in mind. Choosing
a disk in $\mathbb{R}^{2}$ so that one crossing is inside this disk.
If the three links differ by the three crossings $L_{oo},L_{o},L_{\infty}$
(see figure \ref{fig:skein-crossings}) inside of the disk then these
links are skein related. %
Then in $\mathbb{C}\mathcal{L}[[t]]$ one writes the skein relation%
\footnote{The relation depends on the group $SL(2,\mathbb{C})$.%
} $L_{\infty}-tL_{o}-t^{-1}L_{oo}$. Furthermore let $L\sqcup O$ be
the disjoint union of the link with a circle then one writes the framing
relation $L\sqcup O+(t^{2}+t^{-2})L$. Let $S(M)$ be the smallest
submodul of $\mathbb{C}\mathcal{L}[[t]]$ containing both relations,
then we define the Kauffman bracket skein modul by $K_{t}(M)=\mathbb{C}\mathcal{L}[[t]]/S(M)$.
We list the following general results about this modul:
\begin{itemize}
\item The modul $K_{-1}(M)$ for $t=-1$ is a commutative algebra.
\item Let $S$ be a surface then $K_{t}(S\times[0,1])$ caries the structure
of an algebra.
\end{itemize}
The algebra structure of $K_{t}(S\times[0,1])$ can be simple seen
by using the diffeomorphism between the sum $S\times[0,1]\cup_{S}S\times[0,1]$
along $S$ and $S\times[0,1]$. Then the product $ab$ of two elements
$a,b\in K_{t}(S\times[0,1])$ is a link in $S\times[0,1]\cup_{S}S\times[0,1]$
corresponding to a link in $S\times[0,1]$ via the diffeomorphism.
The algebra $K_{t}(S\times[0,1])$ is in general non-commutative for
$t\not=-1$. For the following we will omit the interval $[0,1]$
and denote the skein algebra by $K_{t}(S)$. Furthermore we remark,
that \emph{all results remain true if we use an intersection in $L_{\infty}$
instead of an over- or undercrossing.}

\emph{Ad hoc} the skein algebra is not directly related to the foliation.
We used only the fact that there is an idempotent in the $C^{*}$
algebra represented by a closed curve. It is more satisfying to obtain
a direct relation between both construction. Then the von Neumann
algebra of the foliation is the result of a quantization in the physical
sense.

With more care (see \cite{AsselmKrol2011c}) we can construct a relation
of the skein algebra to the Temperley-Lieb algebra seen as a subalgebra
of the hyperfinite factor $I\! I_{1}$ algebra. Then we have a closed
circle:
\begin{enumerate}
\item The Thurston foliation $F$ of the $S^{3}$ is associated to the hyperfinite
factor $I\! I\! I_{1}$algebra.
\item Using Tomita-Takesaki-theory, one has a continuous decomposition (as
crossed product) of any factor $I\! I\! I$ algebra $M$ into a factor
$I\! I_{\infty}$ algebra $N$ together with a one-parameter group
$\left(\theta_{\lambda}\right)_{\lambda\in\mathbb{R}_{+}^{*}}$ of
automorphisms $\theta_{\lambda}\in Aut(N)$ of $N$. The factor $I\! I_{\infty}$
algebra is isomorphic to $I\! I_{1}\otimes I_{\infty}$.
\item As Jones \cite{Jon:83} showed, the factor $I\! I_{1}$ is given by
an infinite sequence of Temperley-Lieb algebras. 
\item One can construct the foliation $F'$, the horocycle flow of $T_{1}S$
of the surface $S$, with factor $I\! I_{\infty}$algebra related
to $F$.
\item The skein algebra $K_{t}(S)$ represents the horocycle flow foliation
isomorphic to the Temperley-Lieb algebra or the factor $I\! I_{1}$.
\item But the skein algebra is the quantization of a Poisson algebra given
by complex functions over $X(S,SL(2,\mathbb{C}))$. Therefore the
operator algebra of the foliation $F'$ (and also of $F$) comes from
the quantization of a classical Poisson algebra (via deformation quantization).
\item One of the central elements in the algebraic quantum field theory
is the Tomita-Takesaki theory leading to the $I\! I\! I_{1}$ factor
as vacuum sector \cite{Borchers2000}. This factor can be now constructed
by using the foliation induced by an exotic small $\mathbb{R}^{4}$.
\end{enumerate}

\section{Quantum field theory (QFT)\label{sec:Quantum-field-theory}}

In this paper we will understand a relativistic QFT as an algebraic
QFT (AQFT) in the sense of Haag-Kastler (Local quantum field theory)
\cite{Haag:96}. A series of results, accumulated over a period of
more than thirty years, indicates that the local algebras of relativistic
QFT are type $I\! I\! I$ von Neumann algebras, and more specifically,
hyperfinite type$I\! I\! I_{1}$ factors. Therefore our relation between
exotic $\mathbb{R}^{4}$ and this factor is a crucial step in the
understanding of QFT from the geometrical point of view. Any realistic
QFT starts with an action to fix the fields. Here we will discuss
a method to derive this action using the structure of the exotic $\mathbb{R}^{4}$
(see \cite{AsselmeyerRose2010}). 

We will start with some remarks. After the work of Witten \cite{Wit:88.1}
on topological QFTs, there were a growing interest in the relation
between QFT and topology. Witten proposed a supersymmetric QFT to
obtain the Donaldson polynomials as expectation values. Then the further
development of this approach led directly to the Seiberg-Witten invariants
\cite{SeiWit:94,SeiWit:94.2,Wit:94SW}. Therefore one would expect
a strong relation between QFT and 4-dimensional differential topology.
Our approach above gives some hint in this direction. 

The interpretation of this relation is given by the infinite constructions
in 4-manifold theory like Casson handle \cite{Cas:73,Fre:79,Fre:82}
or capped gropes \cite{Fre:83,FreQui:90} used in the construction
of the relation between codimension-1 foliations and exotic $\mathbb{R}^{4}$.
All these structures are described by trees. For example a Casson
handle $CH$ is specified up to (orientation preserving) diffeomorphism
(of pairs) by a labeled finitely-branching tree with base-point {*},
having all edge paths infinitely extendable away from {*}. Each edge
should be given a label $+$ or $-$. Here is the construction: tree
$\to CH$. Each vertex corresponds to a kinky handle; the self-plumbing
number of that kinky handle equals the number of branches leaving
the vertex. The sign on each branch corresponds to the sign of the
associated self plumbing. The whole process generates a tree with
infinite many levels. In principle, every tree with a finite number
of branches per level realizes a corresponding Casson handle. The
Casson handle is at the heart of small exotic $\mathbb{R}^{4}$ (see
\cite{BizGom:96}).

Therefore we have to understand one level of the tree. Then the infinite
repetition of the levels forms the tree. But the basic structure in
QFT is also given by a tree expressing the renormalized fields by
the forest formula \cite{GurauMagnenRivasseau2009}. So we start with
the basic structure of the Casson handle forming the levels of the
tree in 4-manifold theory, the immersed disk. Now we will follow the
paper \cite{Friedrich1998} very closely. Let $D^{2}$ be a 2-disk
and $M$ a 4-manifold as spacetime. The map $i:D^{2}\to M$ is called
an immersion if the differential $di:TD^{2}\to TM$ is injective.
It is known from singularity theory \cite{GuiPol:74} that every map
of a 2-manifold into a 4-manifold can be deformed to an immersion,
the immersion may not be an embedding i.e. the immersed disk may have
self-intersections. For the following discussion we consider the immersion
$D^{2}\to U\subset\mathbb{R}^{4}$ of the disk into one chart $U$
of $M$. 

For simplicity, start with a toy model of an immersion of a surface
into the 3-dimensional Euclidean space. Let $f:M^{2}\to\mathbb{R}^{3}$
be a smooth map of a Riemannian surface with injective differential
$df:TM^{2}\to T\mathbb{R}^{3}$, i.e. an immersion. In the \emph{Weierstrass
representation} one expresses a \emph{conformal minimal} immersion
$f$ in terms of a holomorphic function $g\in\Lambda^{0}$ and a holomorphic
1-form $\mu\in\Lambda^{1,0}$ as the integral\[
f=Re\left(\int(1-g^{2},i(1+g^{2}),2g)\mu\right)\ .\]
 An immersion of $M^{2}$ is conformal if the induced metric $g$
on $M^{2}$ has components\[
g_{zz}=0=g_{\bar{z}\bar{z}}\,,\: g_{z\bar{z}}\not=0\]
and it is minimal if the surface has minimal volume. Now we consider
a spinor bundle $S$ on $M^{2}$ (i.e. $TM^{2}=S\otimes S$ as complex
line bundles) and with the splitting\[
S=S^{+}\oplus S^{-}=\Lambda^{0}\oplus\Lambda^{1,0}\]
Therefore the pair $(g,\mu)$ can be considered as spinor field $\varphi$
on $M^{2}$. Then the Cauchy-Riemann equation for $g$ and $\mu$
is equivalent to the Dirac equation $D\varphi=0$. The generalization
from a conformal minimal immersion to a conformal immersion was done
by many authors (see the references in\cite{Friedrich1998}) to show
that the spinor $\varphi$ now fulfills the Dirac equation\begin{equation}
D\varphi=K\varphi\label{eq:conformal-immersion-Dirac}\end{equation}
where $K$ is the mean curvature (i.e. the trace of the second fundamental
form). The minimal case is equivalent to the vanishing mean curvature
$H=0$ recovering the equation above. Friedrich \cite{Friedrich1998}
uncovered the relation between a spinor $\Phi$ on $\mathbb{R}^{3}$
and the spinor $\varphi=\Phi|_{M^{2}}$: if the spinor $\Phi$ fulfills
the Dirac equation $D\Phi=0$ then the restriction $\varphi=\Phi|_{M^{2}}$
fulfills equation (\ref{eq:conformal-immersion-Dirac}) and $|\varphi|^{2}=const$.
Therefore we obtain\begin{equation}
H=\bar{\varphi}D\varphi\label{eq:mean-curvature-surface}\end{equation}
 with $|\varphi|^{2}=1$. 

Now we will discuss the more complicated case. For that purpose we
consider the kinky handle which can be seen as the image of an immersion
$I:D^{2}\times D^{2}\to\mathbb{R}^{4}$. This map determines a restriction
of the immersion $I|_{\partial}:\partial D^{2}\times D^{2}\to\mathbb{R}^{4}$
with image a knotted solid torus $T(K)=I|_{\partial}(\partial D^{2}\times D^{2})$.
But a knotted solid torus $T(K)=K\times D^{2}$ is uniquely determined
by its boundary $\partial T(K)=K\times\partial D^{2}=K\times S^{1}$,
a knotted torus given as image $\partial T(K)=I_{\partial\times\partial}(T^{2})$
of the immersion $I|_{\partial\times\partial}:T^{2}=S^{1}\times S^{1}\to\mathbb{R}^{3}$.
But as discussed above, this immersion $I|_{\partial\times\partial}$
can be defined by a spinor $\varphi$ on $T^{2}$ fulfilling the Dirac
equation\begin{equation}
D\varphi=H\varphi\label{eq:2D-Dirac}\end{equation}
with $|\varphi|^{2}=1$ (or an arbitrary constant) (see Theorem 1
of \cite{Friedrich1998}). The transition to the case of the immersion
$I|_{\partial}$ can be done by constructing a spinor $\phi$ out
of $\varphi$ which is constant along the normal of the immersed torus
$T^{2}$. As discussed above a spinor bundle over a surface splits
into two sub-bundles $S=S^{+}\oplus S^{-}$ with the corresponding
splitting of the spinor $\varphi$ in components\[
\varphi=\left(\begin{array}{c}
\varphi^{+}\\
\varphi^{-}\end{array}\right)\]
and we have the Dirac equation\[
D\varphi=\left(\begin{array}{cc}
0 & \partial_{z}\\
\partial_{\bar{z}} & 0\end{array}\right)\left(\begin{array}{c}
\varphi^{+}\\
\varphi^{-}\end{array}\right)=H\left(\begin{array}{c}
\varphi^{+}\\
\varphi^{-}\end{array}\right)\]
with respect to the coordinates $(z,\bar{z})$ on $T^{2}$. In dimension
3 we have a spinor bundle of same fiber dimension then the spin bundle
$S$ but without a splitting into two sub-bundles. Now we define the
extended spinor $\phi$ over the solid torus $\partial D^{2}\times D^{2}$
via the restriction $\phi|_{T^{2}}=\varphi$. Then $\phi$ is constant
along the normal vector $\partial_{N}\phi=0$ fulfilling the 3-dimensional
Dirac equation\begin{equation}
D^{3D}\phi=\left(\begin{array}{cc}
\partial_{N} & \partial_{z}\\
\partial_{\bar{z}} & -\partial_{N}\end{array}\right)\phi=H\phi\label{eq:Dirac-equation-3D}\end{equation}
induced from the Dirac equation (\ref{eq:2D-Dirac}) via restriction
and where $|\phi|^{2}=const.$ Especially we obtain for the mean curvature\begin{equation}
H=\bar{\phi}D^{3D}\phi\label{eq:mean-curvature-3D}\end{equation}
of the knotted solid torus $T(K)$ (up to a constant from $|\phi|^{2}$).
Or in local coordinates \begin{equation}
H=\overline{\phi}\sigma^{\mu}D_{\mu}^{3D}\phi\label{eq:trace-second-fund-form}\end{equation}
with the Pauli matrices $\sigma^{\mu}$. Thus the level of the tree
is described by the mean curvature $H$ of the knotted solid torus
$T(K)$ or invariantly by the integral\begin{equation}
\intop_{T(K)}H_{T(K)}\sqrt{g_{\partial}}d^{3}x=\intop_{T(K)}\psi\gamma^{\mu}D_{\mu}\overline{\psi}\sqrt{g_{\partial}}d^{3}x\label{eq:action-3D}\end{equation}
with the metric $g_{\partial}$ at $T(K)$, i.e. by the Dirac action.
Now we will discuss the extension from the 3D to the 4D case. Let
$\iota:D^{2}\times S^{1}\hookrightarrow M$ be an immersion of the
solid torus $\Sigma=D^{2}\times S^{1}$ into the 4-manifold $M$ with
the normal vector $\vec{N}$. The spin bundle $S_{M}$ of the 4-manifold
splits into two sub-bundles $S_{M}^{\pm}$ where one subbundle, say
$S_{M}^{+},$ can be related to the spin bundle $S_{\Sigma}$. Then
the spin bundles are related by $S_{\Sigma}=\iota^{*}S_{M}^{+}$ with
the same relation $\phi=\iota_{*}\Phi$ for the spinors ($\phi\in\Gamma(S_{\Sigma})$
and $\Phi\in\Gamma(S_{M}^{+})$). Let $\nabla_{X}^{M},\nabla_{X}^{\Sigma}$
be the covariant derivatives in the spin bundles along a vector field
$X$ as section of the bundle $T\Sigma$. Then we have the formula\begin{equation}
\nabla_{X}^{M}(\Phi)=\nabla_{X}^{\Sigma}\phi-\frac{1}{2}(\nabla_{X}\vec{N})\cdot\vec{N}\cdot\phi\label{eq:covariant-derivative-immersion}\end{equation}
with the obvious embedding $\phi\mapsto\left(\begin{array}{c}
\phi\\
0\end{array}\right)=\Phi$ of the spinor spaces. The expression $\nabla_{X}\vec{N}$ is the
second fundamental form of the immersion with trace the mean curvature
$2H$. Then from (\ref{eq:covariant-derivative-immersion}) one obtains
a similar relation between the corresponding Dirac operators\begin{equation}
D^{M}\Phi=D^{3D}\phi-H\phi\label{eq:relation-Dirac-3D-4D}\end{equation}
 with the Dirac operator $D^{3D}$ defined via (\ref{eq:Dirac-equation-3D}).
Together with equation (\ref{eq:Dirac-equation-3D}) we obtain\begin{equation}
D^{M}\Phi=0\label{eq:Dirac-equation-4D}\end{equation}
 i.e. $\Phi$ is a parallel spinor. 

\emph{Conclusion:} There is a relation between a 3-dimensional spinor
$\phi$ on a 3-manifold $\Sigma$ fulfilling a Dirac equation $D^{\Sigma}\phi=H\phi$
(determined by the immersion $\Sigma\to M$ into a 4-manifold $M$)
and a 4-dimensional spinor $\Phi$ on a 4-manifold $M$ with fixed
chirality ($\in\Gamma(S_{M}^{+})$ or $\in\Gamma(S_{M}^{-})$) fulfilling
the Dirac equation $D^{M}\Phi=0$. 

From the Dirac equation (\ref{eq:Dirac-equation-4D}) we obtain the
the action\[
\intop_{M}\bar{\Phi}D^{M}\Phi\sqrt{g}\: d^{4}x\]
as an extension of (\ref{eq:action-3D}) to the whole spacetime $M$.
By variation of the action (\ref{eq:action-3D}) we obtain an immersion
of minimal mean curvature, i.e. $H=0$. Then we can identify via relation
(\ref{eq:relation-Dirac-3D-4D}) the 4-dimensional and the 3-dimensional
action via \[
S_{F}(M)=\intop_{M}\bar{\Phi}D^{M}\Phi\sqrt{g_{M}}\: d^{4}x=\intop_{T(K)}\bar{\phi}D^{3D}\phi\:\sqrt{g_{\partial}}\, d^{3}x=\intop_{T(K)}H_{T(K)}\sqrt{g_{\partial}}\, d^{3}x\]
Therefore the 3-dimensional action (\ref{eq:action-3D}) can be extended
to the whole 4-manifold (but for a spinor $\Phi$ of fixed chirality).
Finally we showed that the spinor can be extended to the whole 4-manifold
$M$. 

The integral over the mean curvature $H_{T(K)}$ on the RHS of the
action can be also expressed by\[
\intop_{T(K)}H_{T(K)}\sqrt{g_{\partial}}d^{3}x=\intop_{T(K)}tr(\theta\wedge R)\]
 by using a frame $\theta$ and the curvature 2-form $R$. This term
can be interpreted as the boundary term of the Einstein-Hilbert action.
As shown by York \cite{York1972}, the fixing of the conformal class
of the spatial metric in the ADM formalism leads to a boundary term
which can be also found in the work of Hawking and Gibbons \cite{GibHaw1977}.
Also Ashtekar et.al. \cite{Ashtekar08,Ashtekar08a} discussed the
boundary term in the Palatini formalism. Therefore we have\[
\intop_{T(K)}tr(\theta\wedge R)\to\intop_{M(K)}R_{M}\sqrt{g_{M}}d^{4}x=S_{EH}\]
with $\partial M(K)=T(K)$. Thus we obtain the Einstein-Hilbert action
for the target space of the knotted torus $T(K)$ as representation
of the immersed disk. Up to now we have studied the immersion of the
disk (as well its neighborhood via its boundary the knotted solid
torus $T(K)$) and the relation to the target space. One problem is
left: the connection of the disks to each other. Geometrically we
have a cobordisms between two knotted solid tori $T(K_{n})$ and $T(K_{m})$,
the connecting tube $T(K_{n},K_{m})$. The corresponding integral\[
\intop_{\partial T(K_{n},K_{m})}H_{T(K_{n},K_{m})}\:\sqrt{g}d^{3}x=\intop_{\partial T(K_{n},K_{m})}tr(\theta\wedge F)\]
can be convert to \[
S(\partial T(K_{n},K_{m}))=\intop_{T(K_{n},K_{m})}tr(\tilde{F}\wedge\tilde{F})=\pm\intop_{T(K_{n},K_{m})}tr(\ \tilde{F}\wedge*\tilde{F})\]
by using the Stokes theorem, the extension of the curvature $F$ on
$\partial T(K_{n},K_{m})$ to $\tilde{F}$ on $T(K_{n},K_{m})$ and
the instanton equation $\tilde{F}=\pm*\tilde{F}$ along the tube.
But then we have the Yang-Mills action for the curvature $\tilde{F}$.
We will not discuss the details and refer to\cite{AsselmeyerRose2010}.
Finally we will obtain the correct gauge group $U(1)\times SU(2)\times SU(3)$
and get the action \begin{equation}
S(M)=\intop_{M}\left(R+\sum_{n}(\bar{\Phi}D^{M}\Phi)_{n}\right)\sqrt{g}\: d^{4}x+\sum_{n,m}\intop_{M}tr(\tilde{F}\wedge*\tilde{F})_{nm}\,.\label{eq:action-on-M-standard-model}\end{equation}

As conclusion we can state that an immersed disk used in the construction
of exotic $\mathbb{R}^{4}$ are described by a parallel spinor $\Phi$.
The correspondence goes further because the spinor $\Phi$ as solution
of the Dirac equation (\ref{eq:Dirac-equation-4D}) is not only generated
by a propagator but also by the immersed disk itself. The Feynman
path integral of this action can be rearranged by a simply reorganization
of the perturbative series in terms of trees \cite{RivasseauDisertori2000}.
It should be especially emphasized that this method do not need any
discretization of the phase space or cluster expansion. Then we obtain
a close relation between trees and renormalization similar to approach
of Connes and Kreimer \cite{ConnesKreimer1998}. We close this paper
with these conjectural remarks.

\ack{}{}

We want to acknowledge many fruitful discussion with J. Krol and
H. Rose.

\providecommand{\newblock}{}

\end{document}